# CAN A CYLINDRICAL GEOMETRY DESCRIBE REACTION-DRIVEN DIFFUSION IN NANOMETRIC POROUS MEDIA?

P. C. T. D'Ajello[*], M.L.Sartorelli, M. C. Ribeiro and L. Lauck

Universidade Federal de Santa Catarina

Departamento de Física/CFM

Brazil

P. O. Box 476 - CEP 88040-900

Fax 55 (48) 3721 9946

e-mail: pcesar@fisica.ufsc.br

**ABSTRACT**

We present a theoretical model describing the current transient behavior observed during electrodeposition in an opal modified electrode. As a basic unit we consider a cylindrical vessel with a porous and corrugated surface, whose radius changes periodically with z, the vertical axis of the corrugated vessel. According to the model, the porous network is formed by the replication of those units, put side by side in close contact, and impregnated by an electrolytic solution. Through the lateral surface of those cylinders we allow for a selective flux of species. The inward or outward flux obeys a complex dynamics regulated by the competition between the diffusion kinetics and the chemical kinetics that answer for the reduction of species at a reactive surface located at the bottom of the cylindrical cavities. The analytical expression for the current transient is complemented by a random prescription for the influx or outfluxes of matter through the lateral surface plus a modulation in its intensity that follows the surface corrugation. The theoretical data are compared with the current transients obtained in nanosphere lithography experiments.

## 1. INTRODUCTION

This article presents a theoretical method for describing electrochemical deposition dynamics in nanometric porous media. The nanometric dimensions of the pores will set the special conditions of the system: the effect of capillarity is strong enough to prevent the action of gravitational forces and there is no movement of the fluid as a whole, i.e., there is no drift velocity. The fluid stays in quiescent conditions and the generalized chemical forces are the sole responsible for driving species throughout the liquid medium. This situation is typical in electrochemical deposition through colloidal masks made of nanospheres, to mention an experimental situation that motivates our theoretical work.

In the experimental array whereupon we focus our attention, the porous system is composed of submicrometer spheres ordered in a face centered cubic structure (*fcc*), on top of a flat electroactive substrate [1-3]. The electrolytic medium fills the interstices among the spheres. The geometry dictated by the nanosphere's close packed array, plus a heterogeneous reaction, produce a material deposit that grows on top of the electroactive surface and fills the interstices among the spheres. In [2] it was shown that, experimentally, in order to obtain reproducible current transients in opal modified electrodes it is necessary to work with homogeneously thick colloidal crystals. In Figure 1 we show some examples taken from [2] of experimental current transients that were obtained during cobalt electrodeposition through colloidal polystyrene (PS) masks (the diameters of the PS spheres are indicated in the figure and will correspond to 2R in the present theoretical description). One observes that all transient curves attain a minimum that corresponds to the instant when the metal/liquid interface reaches the equatorial plane of the spheres. Afterwards the current reaches a plateau, when the spheres are completely buried by the metal deposit. One observes that the current plateaus obtained for the 165 nm and 600 nm masks matches the one measured for a plain electrode of same area (compact film). Those masks were nearly perfect monolayers, i.e., the electroactive area of the electrodes were completely covered by a single and ordered array of spheres. The current plateau observed for the 496 nm mask is lower and that is because the mask thickness in that case was not homogeneous, i.e., there were regions covered by a second layer of spheres.

The very nice reproducibility observed in masks of homogeneous thickness is what prompted us to seek for a theoretical description of such system. From a theoretical point of view this class of system (with one, two, three, or n layers), is a prototypical porous medium, where particles, diluted in a quiescent medium, diffuse towards a reactive surface. It is well known that diffusion in porous media is a sluggish process when compared to diffusion in free space. It is generally assumed that this is caused by the limited volume available for diffusion and by the meandering paths of the diffusing particles. To account for that a common approach is to assume an effective medium description, where the diffusion constant is rescaled by the porosity $\varepsilon$ (void fraction) and by the tortuosity $\tau$ (ratio of the effective path length to the shortest path length in the microstructure) that describe the porous medium [1]. In other words, geometry becomes incorporated in the physical description. The regularity of our porous system, however, allows us to decouple geometry from the reaction-diffusion problem. Therefore, in the present model it will be assumed that the diffusion constant is the one that is measured in free diffusion experiments. Moreover, it will be assumed that the diffusivity of species, which according to the Einstein´s equation, depends on the temperature and viscosity of the fluid and on the size of the diffusing particle, will remain constant throughout the experiment.

Briefly stated, by taking these and some other simplifying assumptions that will be explained in the following, we intend to arrive at an analytical expression that describes diffusion driven by reaction in porous systems. Comparison with experimental results will show how far those simplifying assumptions can take us and will point to new directions towards a better description of the phenomenon.

The model tries to capture the essential features of the porous system, reducing it to a problem with cylindrical symmetry, amenable to analytical calculations. The following line of reasoning is considered. A colloidal crystal that is self-ordered on top of a flat substrate has its *(111)* axis directed perpendicular to the substrate, as shown in the top view of Fig. 2a. Along the *(111)* direction there is no straight path to the bottom and the ions are forced to deflect laterally in order to reach down the next layer of spheres. In Fig. 2a the curved white arrows single out one possible diffusion path for an ion that strikes the location marked with a star. The same diffusion path is shown by solid black arrows in the schematic vertical cross section of the porous structure (Fig. 2b). This particular path will be modeled as a staircased cylindrical vessel. At every inflection point along the twisted cylinder the diffusing particle may either remain in the same vessel or diffuse out into one of two other neighbouring units, which are indicated by the dotted white arrows in Fig. 2b. Conversely, at each inflection point a cylinder may receive particles coming from two other vessels. The porous structure, as a whole, can be seen as a periodic replication of a twisted cylinder, conveniently placed along the x and y direction, as pictured in Fig. 2c. As we see, the essence of the lateral diffusion of ions is to allow for exchange of particles among neighbouring vessels and this will be described by a periodic function *g(z)* that modulates, along the length of the cylinder, the intensity of influx/outflux of particles across the walls, plus a random variable that defines the flux direction at each point of exchange. With these provisions two essential features of the problem are taken into account, namely, the lateral diffusion of ions and its periodic and random nature. Therefore, the twisted shape of the cylinders may now be neglected and the problem reduces to an array of porous and straight cylinders directed along the z axis, with a lateral periodicity that mirrors the hexagonal lattice seen at the surface of the porous structure.

The next point of analysis concerns the description of the internal wall of the cylindrical vessel. In real systems the cross-sectional area of a diffusion path has a complex shape. Along the z direction its open portions, where exchange of particles is possible, are periodically intercalated with closed portions of minimal cross-sectional area that are delimited by the closest approximation among three spheres. Moreover, the lateral exchange of particles has an angular dependence with a trigonal symmetry. The model, however, will ignore those details and focus solely on the periodicity of the cross sectional area exhibited along the z direction, assuming for it a cylindrical symmetry, with no angular dependence in the horizontal plane. The internal cross sectional radius of the vessel will vary, in a continuous and periodic fashion, from a minimum value $R_{min}$ up to $R$, which is the external radius of the cylinder, as shown in Fig.

2d. The need of assuming a corrugated internal surface for the cylinders will become more evident later on.

In what follows, we will solve the diffusion problem for one single cylindrical vessel, taken as a minute electrolytic cell, similar to those considered in D´Ajello *et al.* [4-7]. We will generate an analytical expression for the current transient recorded during the heterogeneous nucleation and deposition of ions at the bottom of the cylinder. To solve the equations of such three-dimensional system, a set of boundary and initial conditions will be designed in order to consider the diffusion of particles toward the bottom of the cylindrical cavity. We consider the particles (ions) to be initially homogeneously distributed in a liquid medium that fills the cylinder. The mathematical description of the phenomena offers the opportunity to test several situations concerning the radial influx of matter through the lateral surface of the cavity. In particular, we list some cases of interest involving diffusion through a cylinder:

(i) Cylinder with rigid and impermeable walls: this condition describes the electrodeposition of nanowires in polymeric membranes or in porous alumina, but should also describe, for cylinders with very large radius, a simple electrodeposition process on a flat electrode in a quiescent medium with no convection.

(ii) Cylinders with rigid and semi-permeable walls submitted to a constant influx or outflux of matter: in principle, this condition does not describe any real situation in electrochemistry that we are aware of. However, in a broad context, it could describe, for example, the process of gas exchange that promotes tissue oxygenation in blood capillaries or the transport of neurotransmitters in the brain tissue [8]. In any case, the introduction of wall permeability is the key factor of this model that will allow for the description of more complex systems, like diffusion through a porous matrix.

(iii) Cylinder with rigid and semi-permeable walls with irregular and spatially variable influx of particle: these boundary conditions are a generalization of the former one and the real motivation of the work. It should describe a flux of particles that crosses the walls of the cylinder in a random way, with intensity and direction that change along the vertical axis of the cylinder.

The model assumes the same boundary and initial conditions that are observed experimentally. Initially the vessel is surrounded by a stationary fluid with a homogeneously low concentration of ions. Then, at t=0, a difference of potential is applied at the electrode surface placed at the bottom of the cylinder, and species start to react. As a consequence, a concentration gradient sets up giving rise to a generalized force that guides the ionic species toward the reactive surface. The species flow but the solvent, in which they are diluted, remains under quiescent conditions.

The paper is organized as follows: in Section 2 we introduce the model and the solution obtained for a cylindrical vessel with straight walls. In Section 3 we examine the current transient profiles of a cylindrical cavity with random influx/outflux of species, directed normal to the lateral surface of the cylinder, with a modulated intensity along the z axis. We also discuss the need of introducing corrugated walls in the cylinder in order to reproduce the experimental data. The proposed model is applied to the experimental curves shown in Fig. 1. In Section 4 we present some comments and conclusions.

## 2. THE MODEL

We look for an expression for the current produced by the reduction and deposition of ions at the electrode surface. The electrode surface lies at the bottom of a cylindrical cavity. The cylindrical cavity is filled by an electrolytic solution that contains the ionic species, homogeneously diluted at the initial time. The ionic concentration inside the finite cylindrical region obeys the balance equation

$$\frac{\partial}{\partial t}C(r,\theta,z,t) = D\nabla^2 C(r,\theta,z,t). \qquad (1)$$

In this equation, $C(r,\theta,z,t)$ is the local ionic concentration in the inner part of the cylindrical cavity, $t$ is the time and $D$ is the diffusion coefficient of the solvated ions. The lateral walls of the cylinder allow for matter exchange with neighboring cylinders, with a flux defined by the boundary condition (3) given below. Because the electrode surface is homogeneously reactive only at the bottom of the right circular cavity, we assume rotational symmetry around the vertical axis of the cylinder and Eq. (1) can be simplified to yield the balance equation in cylindrical coordinates

$$\frac{\partial C}{\partial t} = D\left[\frac{\partial^2 C}{\partial r^2} + \frac{1}{r}\frac{\partial C}{\partial r} + \frac{\partial^2 C}{\partial z^2}\right], \qquad (2)$$

where $z$ is the position in the direction normal to the reactive surface, located at $z = 0$.

In Eq. (2) we may observe the absence of electrical contribution as a mechanism to drive the particles. The basic reason to this proposition rests in the fact that chemical generalized forces (that proportional to the concentration gradient) are many decades greater than the electrical forces (proportional to a potential gradient) Such fact is a consequence of the neutrality of the ionic solutions, composed by many different kind of ions. Thus, when a potential difference is set up, with a prescribed intensity, just one species of ions, among many of same signal, is suppressed by the reaction on the electrode surface. Every reaction suppresses an ion, whose concentration we examine, and doing so also suppress a charge. However, the concentration gradient introduced by this suppression computed for the particular species is much higher than the gradient of charges produced by the suppression of one charge unity, because the number of these is huge, once the charge distribution includes all the ions with same signal, not just those that belongs to the observed species.

To solve Eq. (2), we adopt the following initial and boundary conditions:

$$C(r,z,0) = c_b \qquad R \geq r \geq 0, h \geq z \geq 0, \qquad (3a)$$

$$C(r,0,t) = c_b e^{-kt} \qquad R \geq r \geq 0, h \geq z \geq 0, \qquad (3b)$$

$$C(r,h,t) = c_b \qquad R \geq r \geq 0, \forall t \geq 0, \tag{3c}$$

$$\frac{\partial}{\partial r} C(r,z,t)\bigg|_{r=0} = 0 \qquad h \geq z \geq 0, \forall t \geq 0, \tag{3d}$$

$$\frac{\partial}{\partial r} C(r,z,t)\bigg|_{r=R} = -\alpha c_b (1 - e^{-\nu t}) g(z) \qquad h \geq z \geq 0, t \geq 0. \tag{3e}$$

The initial condition (3a) asserts that, before the electric potential is turned on, the ions are uniformly distributed in the region inside the cylinder, and the ionic concentration equals the bulk concentration $c_b$. The time-dependent boundary condition (3b) determines the temporal evolution of the ionic concentration at the electrode surface. This type of time-dependent boundary condition was thoroughly examined in previous articles [4-6]. Briefly, this condition is regulated by the magnitude of $k$, which describes the reactive properties of the electrode. The parameter $k$ describes the rate at which the surface transfers electrons to reduce the ions. Therefore, $k$ is a factor that defines the reaction kinetics. Despite its simplicity, boundary condition (3b) plays a central role in our description because it defines the concentration drop at the electrode where the ions are reduced and withdrawn from the liquid medium. Here we are assuming that when $t \to \infty$ the electrode surface becomes a perfect sink for the particular species. The efficiency of the surface to suppress the species by reactions grows with time. A more general case could be considered if we wrote Eq. (3b) as $C(r,0,t) = (c_b - c_s)e^{-kt} + c_s$, which allows for a limit concentration $c_s$ at the electrode surface. This generalization is perfectly supported by the method we used to solve the mathematical problem (see appendix). It is just to simplify arguments we used $c_s = 0$ here. In our model, if $k = 0$ there is no reaction and there are no concentration changes at the surface ($C = c_b$ at any time). For $k \neq 0$, the concentration at the electrode surface evolves to a stationary value (when $t \to \infty$): a concentration gradient is produced and ions migrate from the bulk of the solution toward the surface. The boundary condition (3c) guarantees that we are working with a cylinder whose length is at least equal to the stationary diffusion layer *h* that defines a distance from the electrode surface, beyond which the ion concentration is assumed to be constant and equal to the bulk concentration $c_b$. Boundary condition (3d) is a consequence of the rotational symmetry of our system. Finally the time dependent boundary condition (3e) determines the evolution of the concentration of ions that flow through the lateral surface of the cavity ($R$ is the outer cylinder radius), in its normal direction. This boundary condition also depends on $z$, i.e., the flux of matter through the cavity's lateral surface can change along its length. This boundary condition contains the parameter $\alpha$, which quantifies the magnitude of matter flux that crosses the cylindrical lateral surface. Its sign determines the direction of matter flow, inward for negative $\alpha$ or outward if $\alpha$ is positive. The time dependent expression contained inside the brackets is a mathematical requirement to guarantee consistency of the boundary and initial conditions. Thus at $t = 0$ there is no flow and the system is characterized by a constant and homogeneous distribution of matter $(C(r,z,t) = c_b)$. When the potential is switched on, the symmetry is broken off. Ions begin to react at the cylinder bottom and a gradient arises to guide the species toward the electrode surface. In addition, the flow across the lateral area of the cylinder obeys a transient rule, quantified by the magnitude of the constant rate $\nu$ that appears in the exponential argument of Eq. (3e). A physical reasoning to justify the temporal dependent term in boundary condition (3e) is that $\nu$ quantifies the time interval elapsed until an external pumping mechanism (osmotic pressure for example) attains its maximum value according to the rule prescribed by $g(z)$ on the lateral surface of the cavity. Regardless of this interpretation, we wish to emphasize the relevance of boundary condition (3e) for our model. It is this condition that sustains the generality of the model, namely, its capacity to reproduce different situations, according to the sign of $\alpha$ and the form of function $g(z)$. Thus a careful choice of parameter $\alpha$ and function $g(z)$ yields a good description of the diffusion phenomenon and heterogeneous reaction in a porous medium.

Boundary condition (3e) defines a flux of species through the lateral surface of the corrugated cylinder. If $\alpha = 0$ or $g(z) = 0$, there is no flux. If $\alpha \neq 0$ and $g(z) \neq 0$ there is a lateral flux whose magnitude is regulated by $g(z)$, which must be choose appropriately in order to produce a very feeble lateral flux if compared to the one produced along the vertical axis. Thus, when $\alpha$ floats this flux works as a localized fluctuation of matter and correlates the concentration of species among neighboring cylinders. This fluctuations must give origin to a dissipation process that abates diffusivity in porous medium, a subject we do not work here, given we work with a single cylindrical cavity.

The calculations are presented in the Appendix. The solution for the current is given by:

$$I(z,t) = -\bar{z}F\frac{D}{h}\pi R^2 c_b f(t), \quad (4)$$

with $f(t)$ given by:

$$f(t) = 1 - e^{-kt} - 2k\sum_{n=1}^{\infty}\frac{e^{-Dw_n^2t} - e^{-kt}}{w_n^2D - k}cos(w_nz)$$

$$+4\frac{D\alpha}{R}\sum_{n=1}^{\infty} g_n w_n\left[-\frac{\left(e^{-Dw_n^2t} - 1\right)}{w_n^2D} + \frac{\left(e^{-Dw_n^2t} - e^{-vt}\right)}{w_n^2D - v}\right]cos(w_nz), \quad (5)$$

where the following parameters are used:

$$w_n = \frac{n\pi}{h} \quad (6)$$

$$g_n = \int_0^h g(z)\,sin(w_nz). \quad (7)$$

To support its general character we do not specify the form of the function $g(z)$ from the beginning.

## 3. RESULTS

In this section we examine the behavior of the time function $f(t)$. We first consider a cylindrical cavity with flat ($g(z) = \beta = constant$) and impermeable walls $(\alpha = 0)$ to examine the role played by the diffusion coefficient $D$ (Fig. 3a, with $k = 0.89s^{-1}$) and reaction rate constant $k$ (Fig. 3b, with $D = 1.x10^{-6}cm^2s^{-1}$) when all other parameters are kept constant ($R = 300nm,\ h = 3x10^{-3}cm$). Fig. 3a shows that the slower the ionic diffusion, the lower the transient peak and also the stationary current when $t \to \infty$. Fig. 3b shows that increasing the reaction rate k also increases the transient peak, but decreases the time necessary to reach equilibrium. In Figs. 3c-d we consider the effect of permeable walls with constant influx/outflux of particles ($\alpha = const. \neq 0, D = 1.x10^{-6}cm^2s^{-1}$, $k = 0.89s^{-1}$). In this case it is also necessary to define a value for the rate v at which the pumping rate for influx/outflux of matter reaches a maximum. In Fig. 3c, $\nu$ is kept constant at $0.1s^{-1}$, whereas $\alpha$ varies from influx ($negative\,\alpha$) to outflux ($positive\,\alpha$) of matter. It is observed that allowing for ionic permeation through the walls of the cylinder solely modifies the plateau value of $f(t)$, whose absolute value either increases or decreases, depending whether $\alpha$ is negative or positive . As it would be expected, the increase in rate $\nu$ (Fig. 3d) accelerates the time necessary to attain equilibrium.

In a next step we will try to mimic the periodic sequence of points of exchange observed in an ordered porous medium. In order to describe the radial flux of matter (species), the following non-constant $g(z)$ function, as depicted in Fig. 2d, will be assumed:

$$g(z) = \beta cos^2\left(\frac{\pi}{2R}z\right). \quad (8)$$

Expression (8) defines a periodic behavior for the flux of species that enter or leave the cylindrical cavity according to its position in space. There is still a rotational symmetry and also a periodicity, once $g(z) = 0$ for $z = qR$ with $q = 1,3,5 \ldots$ and $g(z) = \beta$ when $= 0,2,4, \ldots.$ . One should note that we still have a cylindrical cavity with flat walls, although the flux of species obeys a spatial distribution that identifies the regions where species have easy access to the cylindrical cavity (the pore). However, it is also necessary to consider the fact that in each pore the diffusing particle may choose either to remain in the same vessel or move to a neighboring one. To describe this complex dynamic behavior of a real porous system, where influx or outflux of particles is at random, we adopt the simplest proposition. We assume that the sign and the magnitude of $\alpha$ are random variables with a null ensemble average, assigning to $\alpha$ a plain random walk behavior. Thus, the change in flow´s intensity and direction through the lateral surface of the cavity is represented by the product of functions $\alpha g(z)$, that defines the representative horizontal bars appearing on Fig. 2d. The random variation in $\alpha$ is achieved by a Monte Carlo algorithm introduced in Eq. (5) considering the fact that, at each layer, the particle may stay in the same vessel or diffuse out into one of two neighboring vessels. When the particle takes one particular path, it is in fact withdrawn from the other two, a situation that undergoes inversion in a future time interval. The following prescription is taken: the initial value is $\alpha_0 = 0$ then, at a certain instant $t_1$ a constant increment $\pm\Delta\alpha$ is chosen with the signal taken at random, which defines a new value for $\alpha$, $\alpha_1 = \alpha_0 \pm \Delta\alpha$, that is maintained for a time interval $\Delta\tau$. The procedure is repeated until the current transients are concluded.

There is a final aspect to consider before the application of the method. As the heterogeneous reaction that occurs at the electrode results in a material deposit at the bottom, there will be a progressive increase in the deposited film thickness and consequently a progressive shift on the *g(z)* function that mimics the lateral flux. This effect changes the flux geometry near the topmost deposited layer. To account for that, the reactive surface is kept fixed at $z = 0$ whereas $z$ is shifted by a discrete and constant quantity, so that we have $g(z')$ with $z' = z + \Delta z$ where $\Delta z$ is the increment on the deposit thickness after a given time interval $\Delta t$. This trick allows us to compute Eq. (5) at $z = 0$ taking into account the correction in $g_n(z)$.

It is also necessary to consider that in experiments with opal modified electrodes the macroscopic electroactive area is kept constant, whereas the density of pores changes accordingly to the diameter of the spheres. For comparisons with experimental results and also among transients of different radii one needs to normalize the theoretical curve by the solid area occupied by the projection of a single vessel.

Fig. 4a presents the current transients obtained for a cylinder with flat walls and a periodically changing permeabilitity with random influx/outflux of material. Eq. (4) is used to generate current transient profiles, using the standard values $D = 1x10^{-6} cm^2 s^{-1}$ and $= 0.89 s^{-1}$ . We also assume that the cylinder radius is $R = 300nm$ and the magnitude of the depletion layer $h$, i.e., the perpendicular distance from the reactive surface, beyond which the concentration of species remains constant and equal to the initial concentration $c_b$ is $h = 3x10^{-3} cm$. Each curve in Fig. 4a corresponds to a different seed for the random number generator, so that there is a different sequence of random numbers that define the signal and magnitude of $\alpha$ in Eq. (5). The instabilities observed in the current transients reflect the random nature of α. The average of the four curves yields a flat curve similar to that drawn by a continuous line in Fig. 1. It is easy to verify that the current has an undefined pattern, that do not resemble the ones (drawn with symbols) shown in Fig. 1, and if we add all of them we get the flat curve, characteristic of a compact film deposition. This result indicates that the modulation of the flux intensity, when combined with a random parameter, which could be thought as the basic ingredient that defines a porous matrix, is not enough to represent it. It is necessary to go further and consider a variation in the cross-sectional area of the cylinders in order to reproduce the pores in a realistic way.

In what follows we will add a periodic corrugation at the lateral walls of the cylinder in order to try to mimic the porous geometry in an effective way. To implement this idea we consider that $\pi R^2$ the cross sectional area of the cavity, is now a periodic function of $z$, described by $\pi \tilde{R}^2$ , as sketched in Fig. 2d.

$$\pi\tilde{R}^2 = \pi\left(R - 0.25R\left[1 - cos^2\left(\frac{\pi}{2R}z\right)\right]\right)^2. \qquad (9)$$

The function $g(z)$ has still the form given by Eq. (8) and $\alpha$ obeys the same prescription as before. Figure 4b presents the theoretical results for a $2R = 600\ nm$ unit cellular cavity, assuming that:

$$g(z) = \beta cos^2\left(\frac{\pi}{2R}z\right) \qquad 0 \le z \le 2R \quad with\ \pi\tilde{R}^2\ given\ by\ Eq.(9).$$

$$g(z) = 0 \qquad z > 2R.$$

The curves described by symbols correspond to current transients obtained from Eq. (4), using the same values applied to generate the curves of Fig. 4a. The continuous black curve is an average of the four curves. It is observed that the current now displays a broad minimum before reaching the stationary value. The results shown in Fig. 4 demonstrate that corrugations along the diffusing path are indeed necessary to describe the characteristic features observed in the experimental current transients.

In Fig. 5 we show three simulated transients corresponding to unitary cells that differ only by the magnitude of $R$. The curves were obtained with the same seed used to generate the curve drawn with open circles in Fig. 4b. A comparison between Figure 5 and Figure 1 shows that the model is able to describe the experiment reasonably well. Particularly this exercise emphasizes the role played by geometry to define the currents, but it also shows there is still room for improvement.

First, one observes in Fig. 1 that the absolute value of the current measured at its point of minimum increases with the radius of the spheres, a fact not verified in theoretical realizations. This experimental evidence indicates that geometry is not the sole factor that regulates the current transient. There is a physical process, not accounted for by the model, which affects differently porous systems of different sizes.

Second, the model is not able to predict correctly what happens when a second layer of spheres is added on top of the first one. That is not shown here but our theoretical results indicate that in current transients obtained for one and two layers of spheres, the first minimum in both curves occurs at the same time. However it was already demonstrated [2], for systems made with 600 nm spheres, that the presence of just one additional layer reduces the amount of deposited material to just 8 % of what is deposited, in the same period, through a single monolayer. This is a very strong reduction in the rate of deposition that is not explained neither in this model or elsewhere [1].

In conclusion, we have developed a general mathematical model that is able to describe the complex interplay between diffusion and reaction in a nanometric cylinder. The model is capable of describing situations where the reaction rate is faster than diffusion, i.e., when deposition is controlled by diffusion. Furthermore, the ability to tailor the internal geometry of the side walls, as well as their degree of permeability and selectivity, offers a powerful tool to play with. Although the pumping mechanism responsible for the influx or outflux of a particular species has not been specified, the regular diffusion of species and the heterogeneous reaction that occurs inside the cylindrical cavity are precisely defined by simple parameters whose meaning are simple and intuitive. In particular, the present model could describe the current transients observed during electrodeposition in porous scaffolds [2, 3]. In this case, however, it is important to bear in mind that the curves shown in Fig. 4 and Fig. 5 represent just one vessel, whereas the experimental current transients represent the average behavior over $10^8 - 10^9$ vessels placed side by side. Furthermore, when dealing with systems composed of many layers of spheres, the correlation length among vessels should become an important issue [3]. The vessels are able to exchange particles with two other vessels in each level, establishing a braided type of network. As time evolves, the correlation among vessels increases. Even when $\alpha$ is a random variable with an ensemble average equal to zero, it depicts a square deviation that is, on the average, different from zero, and this will influence the final result when the statistical computation of all the units that form the system are adequately considered.

## 4. COMMENTS AND CONCLUSIONS

In this paper we demonstrated that an ordered porous system can be represented by a set of cylindrical vessels with permeable walls. A set of boundary and initial conditions for diffusion and reaction in a cylindrical vessel results in an analytical expression for the current transient measured at the bottom of the cylinder that is completely defined by a group of parameters that control the intensity of ionic flux through the walls and the rate of ion consumption at the bottom of the cylinder. It was shown that the introduction of a periodic corrugation in the cylindrical wall is essential to reproduce the current minima observed in experiments. It was also shown that for very large systems, the permeabilitiy of walls may be neglected. Results indicate, however, that despite the good qualitative agreement obtained with experimental data, the model needs to be improved in order to explain some features observed in real systems.

Acknowledgements

The authors would like to acknowledge to the Brazilian agency CNPq, for financial support, and in particular, the INCT of Organic Electronics 573762/2008-2, 380200/2010-4, 4808877/2008-4).

**Figure Captions**

**Figure 1:** Current transients obtained during cobalt electrodeposition through monolayered polystyrene colloidal masks, after [2]. The diameters of the polystyrene spheres are indicated in the figure. The solid line represents the current transient measured at a flat electrode.

**Figure 2:** Can a cylindrical geometry describe an ordered porous media? (a) Top view of a colloidal crystal formed by four layers of spheres self-ordered on a flat substrate in afcc structure. The star at the center indicates one pore through which an impinging ion may enter the porous structure. The white arrows depict one possible diffusion path towards the flat substrate. (b) Schematic cross section of the colloidal crystal. Black arrows indicate the diffusion path that was singled out in (a). (c) That particular diffusion path is now modeled as a staircased cylinder. The whole porous structure can be seen as a periodic replication of aligned twisted vessels. At each inflection point the vessels may exchange particles with neighboring ones. (d) In the last simplification step the twisted vessels transform into straight cylinders with corrugated walls. The function $g(z)$ is sketched on the right, indicating the points of maximum and minimum flux among vessels.

**Figure 3:** Function $f(t)$ given by eq. (5), which represents the temporal part of the current transients. Here $g(z) = \beta = 1.$, which means the absence of corrugation on the cylindrical surface. $c_b = 26\ mM$, $h = 3x10^{-3}cm$ and $v = 0.1\,s^{-1}$. Fig. 3a shows the results for an impermeable cylinder ($\alpha = 0$), with different diffusivity. Fig. 3b shows the function $f(t)$ when using different values for the reaction rate $k$ ($D = 1x10^{-6}cm^2s^{-1}, \beta = 0.0025, \alpha = 0.$). Figures 3c and 3d shows the function $f(t)$ constant ouflux and influx of matter respectively, see the values for $\alpha$ in the figure $(D = 1x10^{-6}cm^2s^{-1}, v = 0.1s^{-1}, k = 0.89s^{-1}\ and\ \beta = 1.)$.

**Figure 4:** Total current flowing through the reactive base of the cylinder. The curves where obtained from Eq.(4) assuming $D = 1x10^{-6}cm^2s^{-1}, k = 0.89s^{-1}, v = 0.1s^{-1}, \beta = 150$. $\alpha$ is a random parameter and in Fig. 4a the cylindrical cavity is not a corrugated one whereas, in Fig. 4b, we have a corrugated cylinder.

**Figure 5:** Theoretical current transients obtained from Eq. (4). The profiles describe diffusion in one corrugated vessel with different radii, as indicated in the figure. The current transients where generated by one and the same seed. All cases show a current that flows through equal reactive area. We used $D = 1.x10^{-6}cm^2s^{-1},\ v = 0.1s^{-1}, \beta = 150., h = 3x10^{-3}s^{-1}\ and\ k = 0.89s^{-1}$.

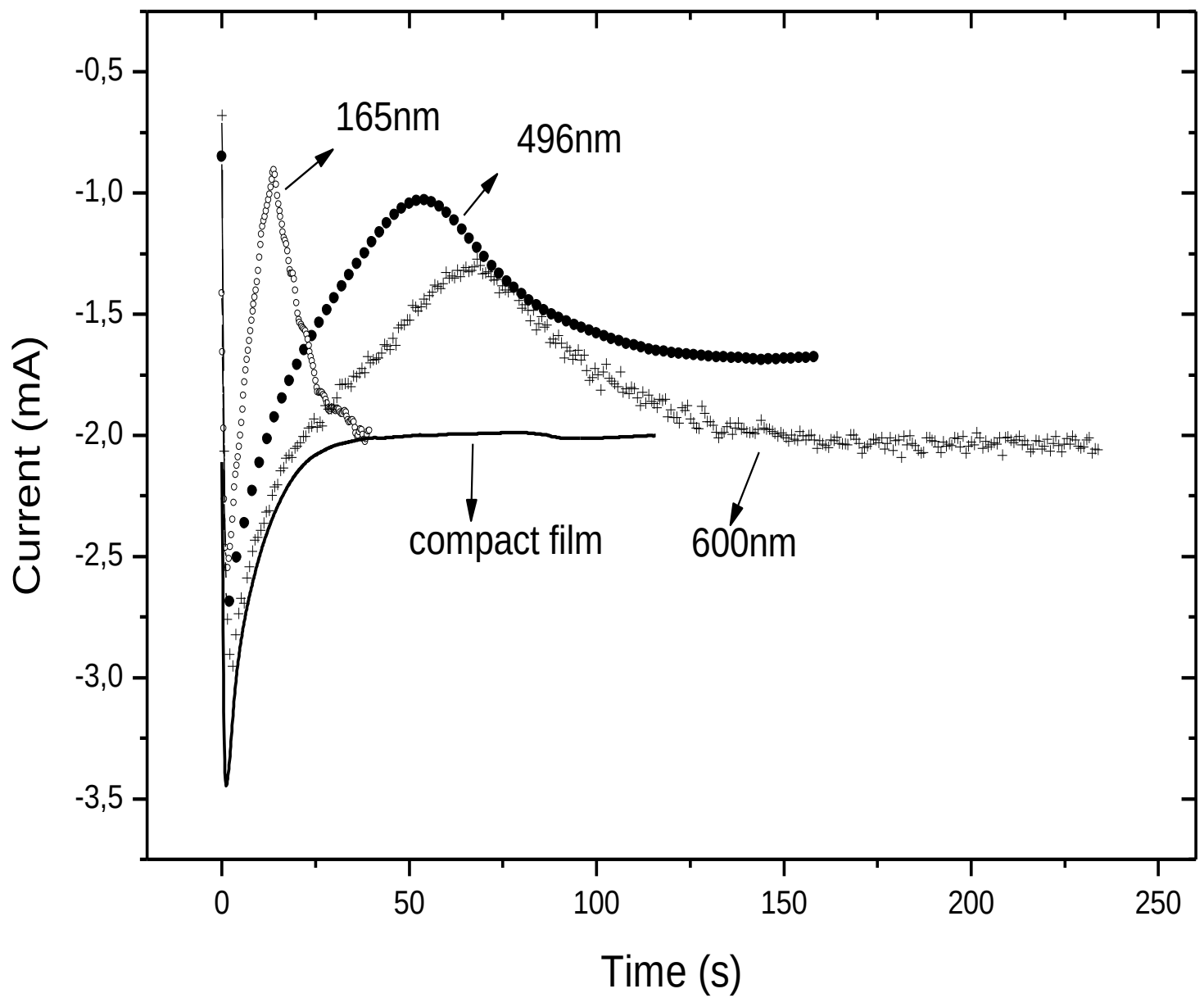

Figure: 1, PCT D´Ajello;

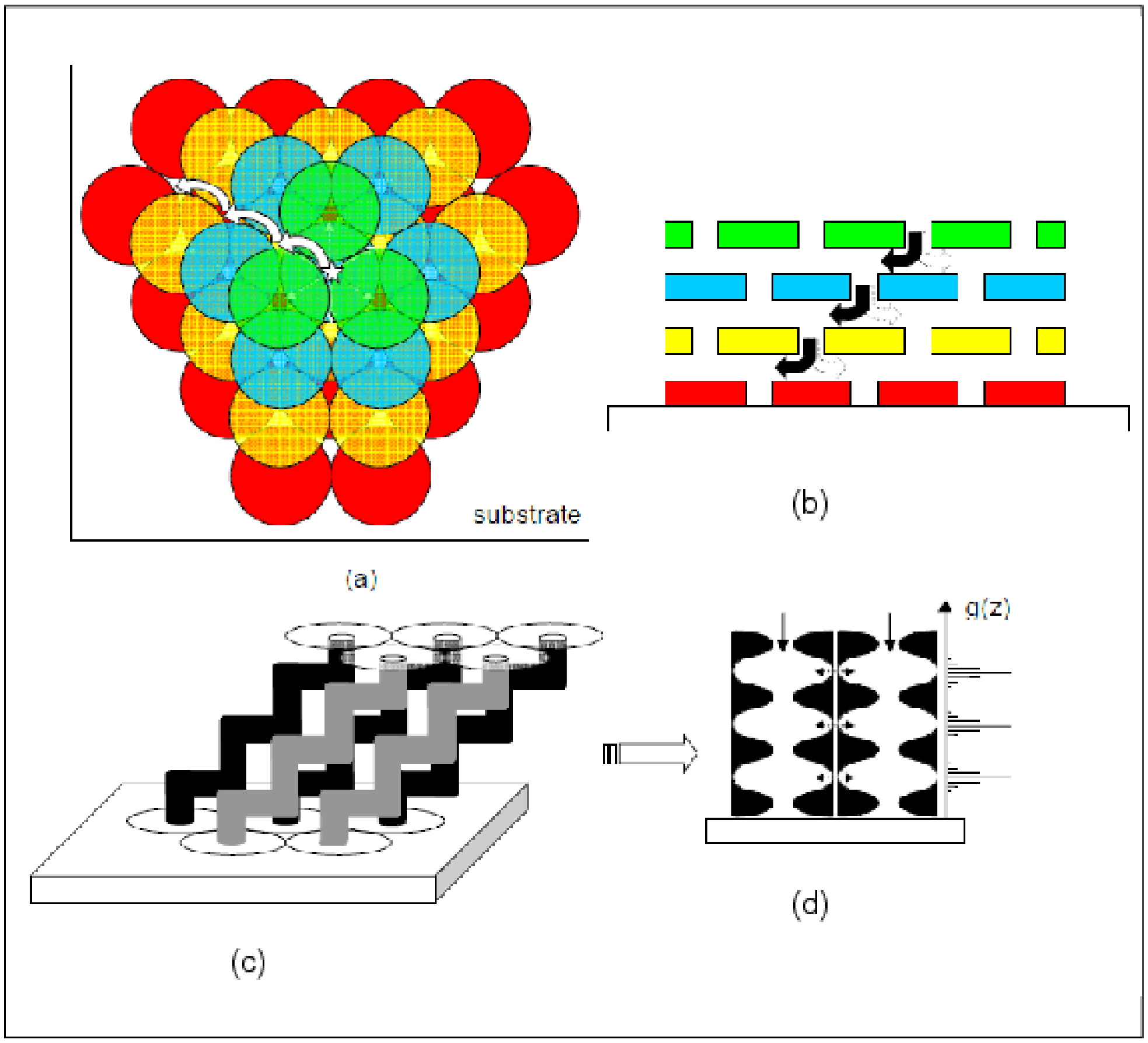

Figure: 2; PCTD´Ajello

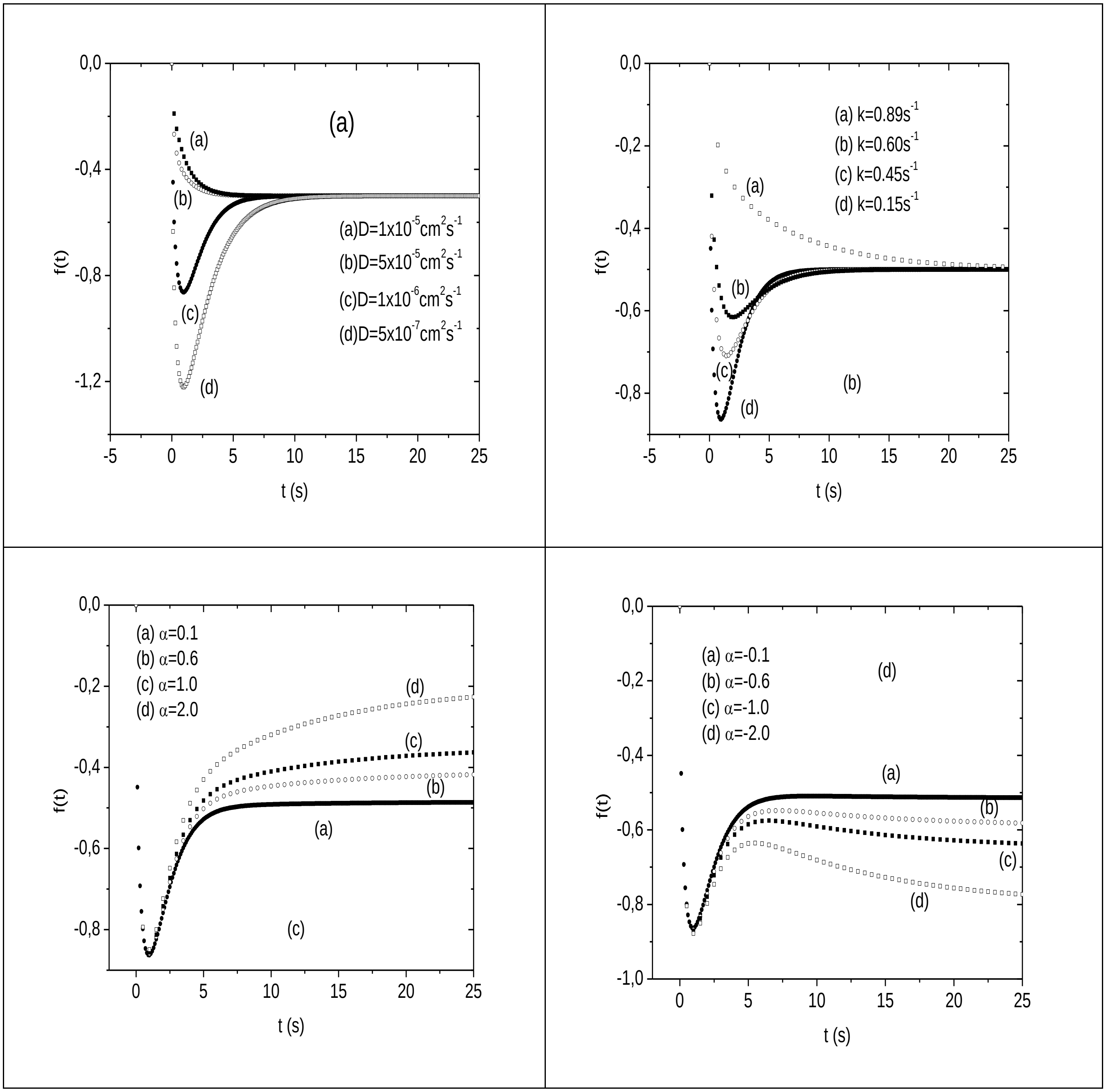

Figure: 3; PCTD´Ajello

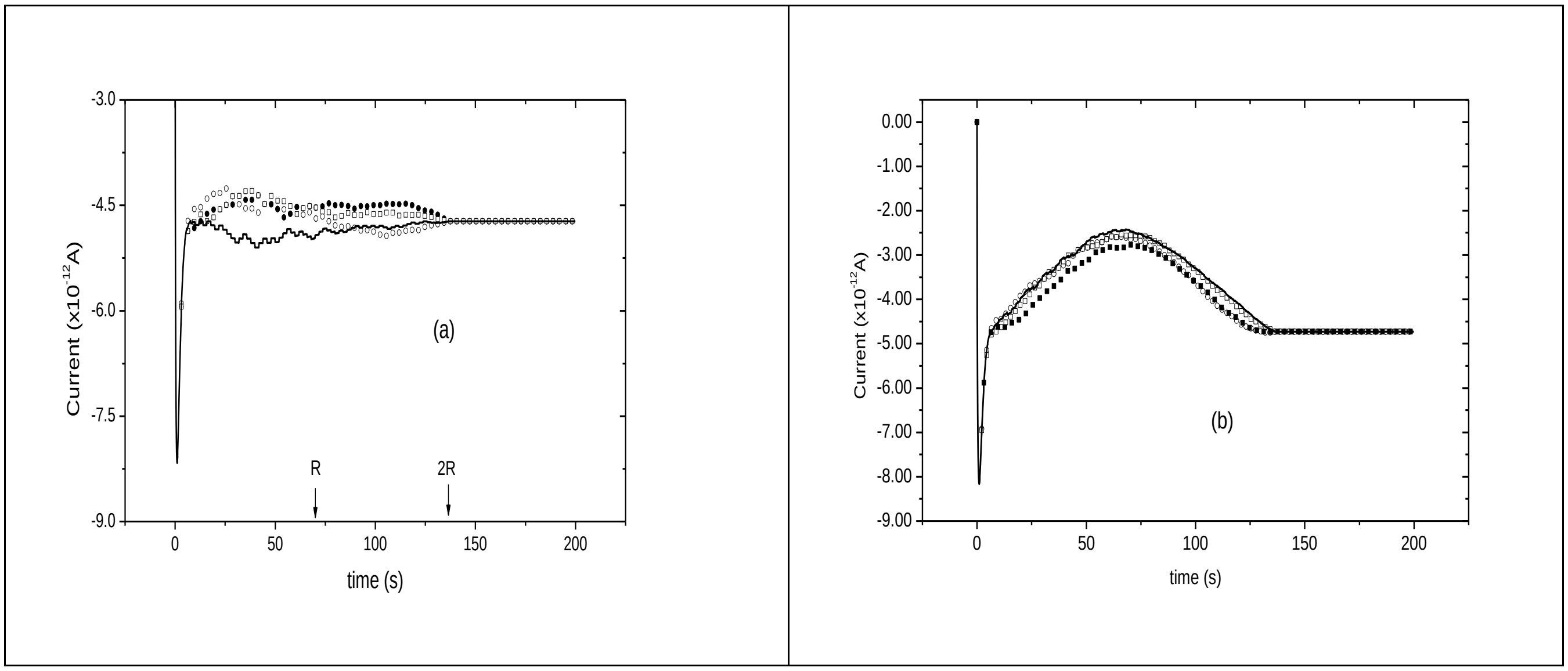

Figure: 4; PCTD´Ajello

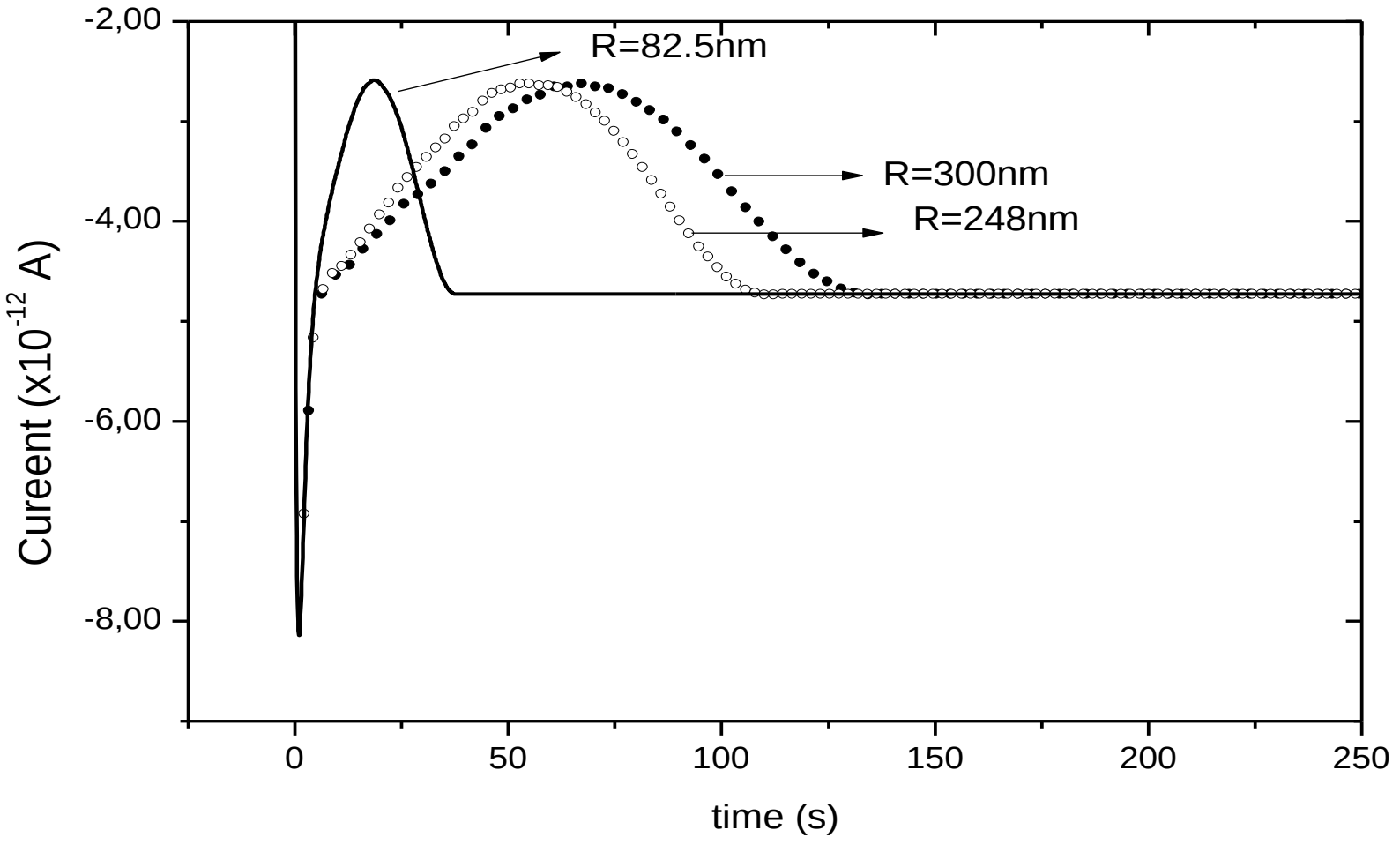

Figure: 5; PCTD´Ajello:

*(d)*